# Emission dynamics in zincblende $InAs_xP_{1-x}$ quantum dots in InP nanowires: influence of quantum dot size, composition and nanowire geometry

Tomasz Gzyl[1], Giada Bucci[2], Krzysztof Gawarecki[3], Elisa García-Tabarés[4], Anna Musiał[1], Valentina Zannier[2], Ylea Vlamidis[2], Fabio Beltram[2], Julian V. Montero[4], Beatriz Galiana[4], Lucia Sorba[2], Wojciech Rudno-Rudziński[1], Grzegorz Sęk[1]

*[1]Department of Experimental Physics, Faculty of Fundamental Problems of Technology, Wrocław University of Science and Technology, 50-370 Wrocław, Poland*

*[2]NEST Istituto Nanoscienze CNR and Scuola Normale Superiore, 56127 Pisa, Italy*

*[3]Institute of Theoretical Physics, Faculty of Fundamental Problems of Technology, Wrocław University of Science and Technology, 50-370 Wrocław, Poland*

*[4]Universidad Carlos III de Madrid, 28903 Leganés, Spain*

*Corresponding author: tomasz.gzyl@pwr.edu.pl*

**Abstract**

Hereby, we present an experimental and theoretical investigation of emission dynamics in zincblende $InAs_xP_{1-x}$ quantum dots (QDs) embedded in InP nanowires (NWs) grown via vapour-liquid-solid mechanism by chemical beam epitaxy, using Au nanoparticles as a nucleation catalyst. By measuring time-resolved photoluminescence from an ensemble of QD-NWs it was possible to determine the exciton lifetime dependence on QD composition and height. Changes in the InP shell thickness surrounding the InP NW stem with a QD, brought additional insight into the influence of photonic environment on the carrier dynamics. High-resolution transmission electron microscopy, combined with energy-dispersive X-ray spectroscopy, provided actual structural parameters. The experimentally obtained lifetimes were interpreted in the light of results of 8 band k·p calculations combined with configuration-interaction model to take into account the Coulomb interactions and finite-difference time domain photonic simulations to include the effect of optical confinement. The full understanding of the experimental results required considering both, the changes in the QD potential and the Purcell effect, the latter leading to spontaneous emission inhibition in the case of NWs with thin InP shell.

## 1 Introduction

One of the key building blocks of quantum information processing and quantum communication is a high-quality single-photon source [1], [2]. An ideal emitter for this purpose has to generate single photons on-demand with high rate and for some applications also exhibit indistinguishability required e.g. in quantum repeaters [3], [4]. Sources emitting in the telecom C or O bands are especially desired, since in these spectral ranges light dispersion and attenuation in silica optical fibres, on which the current global network infrastructure is based, are the lowest. Direct emission at the telecom wavelengths enables efficient realization of quantum communication protocols over large distances, [5], [6] without the need for frequency conversion, which is known to lower the system brightness and to increase the noise level [7], [8].

The best platform that fulfills the abovementioned requirements are epitaxially grown group III-V semiconductor quantum dots (QDs), which exhibit tunable bright and deterministic emission of high purity single photons. They are marked by long-term stability, fabrication scalability, and good integration with mature on-chip semiconductor technology. Single photon emission at the telecom wavelengths has been demonstrated in InAs/InP, InGaAs/GaAs and (In)GaSb/AlGaSb [9], [10] QDs, with emission and extraction efficiency (EE) boosted by placing them in photonic structures, such as mesas, (micro)lenses or various cavities, for instance circular Bragg grating (CBG) microcavities [11].

Another recently extensively developed nanostructure, with promising performance in the telecom spectral range, is an $InAs_xP_{1-x}$ QD embedded in an InP nanowire (QD-NW). The applied growth methods allow for good control over NW geometry, as well as the QD size and composition. What is more, QD position along the NW vertical axis can also be easily tailored, which opens up the possibility to fabricate more complex structures, such as stacks of multiple QDs inside one NW, with fully controlled strength of the interdot coupling between the individual dots [12]. Most commonly, $InAs_xP_{1-x}$ QD-NWs are grown using selective area epitaxy in a vapour-liquid-solid mode (SAE-VLS) on InP substrates, with lithographically fabricated oxide masks [13]. The growth is realized along the (111) crystallographic direction, which results in QD-NWs in the wurtzite crystal phase (WZ). While the technology is very well established for QD-NWs emitting at wavelengths below 1000 nm [14], [15], [16], [17], sustaining high performance at the telecom wavelengths is more challenging. In the WZ phase, catalyst nanoparticles with diameters of 20 nm or less are required to ensure a high crystal quality, which leads to fabrication of ultrathin QD-NWs. This constraint makes it harder to manipulate the QD parameters to tune its emission to longer wavelengths. Nevertheless, in recent works, O- and C-band emissions have been demonstrated with an InAsP QD in an InP NW [18], [19] and an InAsP QD additionally embedded in an InAsP rod with lower As content [20], respectively. The latter enabled observation of single-photon emission with probability of multiphoton emission below 10%.

An alternative and novel solution, which is studied in this work, is a QD-NW of the same material system but grown in the zincblende (ZB) phase along the <100> direction, using chemical beam epitaxy (CBE) technique in a VLS mode. As shown in our previous articles, the ZB QD-NW emission can be tuned to both the telecom O- and C-band by adjusting the QD height and composition [21], [22]. Reaching longer wavelengths is easier in this case because InAs and InP in the ZB phase have smaller energy bandgaps than in the WZ one and the colloid nanoparticle size constraint is lifted for the ZB NW growth as the ZB crystal structure is the only stable configuration, so the defect formation energy is higher compared to the WZ NWs. With this growth method no lithographically prefabricated substrate is used and the control of the NW shell geometry, crucial for high photon EE from the nanostructure, is achieved by balancing the radial and vertical NW growth through changing growth parameters [22].

By proper tailoring of the NW shell thickness (nanowire in-plane dimension) QD spontaneous emission can be almost fully coupled to a single optical mode propagating along the NW vertical axis. The β factor, defined as $\Gamma_{HE11}/(\Gamma_{HE11}+\gamma)$, where $\Gamma_{HE11}$ is the spontaneous emission rate into the fundamental $HE_{11}$ waveguide mode and γ represents emission rate to all the other modes, can reach values close to unity [15]. The waveguiding effect results in an increased photon EE upwards and reduction of radiation losses in radial directions through the dielectric-screening effect [23]. The small NW taper angle (on the order of single degrees) ensures adiabatic mode expansion, efficiently outcoupling emission from the NW tip-air interface, and directional emission with Gaussian far field profile, which favors coupling to external optics and optical fibres with high extraction efficiencies. Additionally, in contrast to demanding planar positioning and spectral matching processes of quantum dots in cavities, QDs in NW are guaranteed to be positioned on the NW vertical axis, which is the optimal spatial position for waveguide mode coupling [24]. Spectrally, the NW platform offers broadband optical response with the bandwidth of tens to hundreds of nanometers, combined with broad and well-controlled emitter spectral tunability [15], [23], [25]. In case of microcavities, this task is by no means trivial, because reaching high quality factors and boosting EE requires precise spatial and subnanometer spectral tuning of the emitter to match the cavity resonance [26]. In order to increase the bandwidth, a lower cavity finesse is compensated by decreasing the optical mode volume. Then, spatial positioning becomes critical and hard to achieve even with deterministic positioning methods, which accuracy reaches values of $15\pm11$ nm (system response corrected value) using advanced imaging techniques [27], [28] and needs to take into account the QD localization uncertainty on the order of <10 nm. What is more, EE from the cavity is further limited by fabrication imperfections and absorption losses [29].

To assess the actual application potential of a structure as a single photon source, it is crucial to obtain knowledge on its emission dynamics. The carrier lifetime influences the measured Hong-Ou-Mandel visibility due to time jitter and photon indistinguishability via its relation to the pure dephasing and decoherence times, as well as imposes fundamental limitation on the maximal photon generation rate, which translates to the efficiency of quantum communication and secret key rates in quantum cryptography. To improve these application-relevant figures of merit, the radiative lifetime can be shortened through Purcell effect coming from the photonic environment. In the case of optimized QD-NW structure, the Purcell factor is typically around 1 [22], [25], but it was shown in the literature that the calculated emission intensity (due to both Purcell enhancement and increase in photon extraction efficiency) can be increased by a factor of 2.7 by implementing a bottom metallic mirror [14]. On the other hand, an insufficiently thick NW shell for guided mode coupling leads to a strong QD spontaneous emission rate inhibition [15], [23]. The elongation of exciton lifetime in thin QD-NWs was confirmed experimentally for WZ single InAsP/InP QD-NWs emitting at 980 nm [30] and for an InAs QDs layer embedded in the GaAs NW emitting at 920 nm [23]. Another factor influencing the radiative lifetime is the electron-hole wavefunctions overlap, which depends on the structural details of the emitter exclusively. In the more common WZ QD-NW system grown by SAE-VLS, the measured lifetime values reported in the literature are in the range of 2.1-2.6 ns for structures emitting in the telecom O-band and 2.2 ns for the telecom C-band [18], [19], [20], [31]. In the ZB case, the structure symmetry is higher than in the WZ one, which reduces the built-in piezoelectric fields being a reason for the spatial electron-hole separation and exciton lifetime elongation in the case of WZ structures [32],[33]. On the other hand, changing the crystallographic structure alters the bands splitting and transition probabilities, effectively changing the exciton lifetimes. Therefore, it is not straightforward to predict the influence of a different crystal structure on the carrier's lifetime, so it needs to be thoroughly studied for each particular case.

In this work, we present comprehensive experimental and theoretical considerations for the ZB InAsP/InP QD-NW system, where we study the effects of QDs size and composition, as well as the NW shell thickness, on the radiative lifetime in the spectral range of 1000-1600 nm. We use (scanning) transmission electron microscopy – (S)TEM – and energy dispersive spectroscopy (EDS) to obtain structural data on geometry, composition and structural quality of our samples, which is then used for theoretical multiband $k \cdot p$ calculations of the QD, together with configuration-interaction model to include the excitonic effects, and for photonic simulations of the QD-NW system to calculate the Purcell enhancement factor ($F_P$). Optical measurements are performed by means of photoluminescence (PL) to study the emission of QD-NW ensembles. Time-correlated single-photon counting technique is employed to obtain PL decay times, giving insight into dynamics of the investigated system.

## 2 Methods

*2.1 Growth*

The zincblende $InAs_xP_{1-x}$ QDs in InP NWs are grown epitaxially using the VLS growth mode in a Riber C-21 CBE system. The growth details are described in our previous work [21], thus the methodology will be summarized here briefly. The growth is initialized by casting commercial water solution of Au colloids (BBInternational EM.GCnn), with a nominal nanoparticle size of 30 nm, onto the Fe-doped InP (100) substrate. The nanoparticle serves as a nucleation center, from which axial growth of the NW begins in the presence of gaseous metalorganic trimethylindium (TMIn) and tert-butyl phosphine (TBP) precursors fluxes. The growth of the QD segment is realized by introducing tertbutyl arsine (TBAs) into the chamber, preceded by a growth interruption. During the QD growth stage, a thin InAsP 2D layer is also deposited on the substrate surface and on the NW sidewalls around the QD, resulting in an InAsP shell with a thickness of a few atomic layers (see Fig.2). After the QD segment is completed, TBAs supply closes and the InP tip is grown to finalize the NW stem. In the next stage, the NW InP tapered shell is grown through adjusting the growth parameters to balance the interplay between the radial and axial NW growth [22].

The final NW morphology of the grown samples is determined by collecting images from the top and 45° tilted view by means of a Zeiss Merlin SEM operating at 5 kV. Based on the images, the NW geometry parameters are measured using image analysis software Image J. These are further used in photonic calculations (Section 3.3) to determine the extraction efficiency and Purcell factor. The QD structural analysis is performed using advanced high-resolution scanning transmission electron microscope (HR-STEM) Thermo Fisher Spectra 200-COLDFEG, with aberration corrected condenser lens. The system provides a resolution of 0.6 Å at 200 kV. In order to determine the QD geometry, size, composition and atomic distribution, dual energy dispersive spectroscopy detector with 1.7 Sr effective solid angle is used, combined with volumetric signal simulation. These results guide the choice of parameters for the calculations of electronic structure (Section 3.2). The details of all the studied samples are presented in Table1. To study the influence of QD size, samples differing in QDs height (size along the NW axis – the direction of the strongest quantization) have been grown – QDs in sample A have 10 nm and in sample B and C – 2 nm. To keep the emission wavelength away from the emission of the 2D InAsP layer on the substrate, the high As content in the QD material has been selected and it is in a similar range for all the samples – from 69% to 76% (based on the structural characterization – see Section 3.1), with sample A having slightly lower As content than samples B and C.

To get an insight into the influence of the photonic environment on the carrier dynamics and learn about the Purcell effect in the investigated structures two NW geometries have been realized – with thin InP shell of 55 nm (sample B) and with thick InP shell of 150 nm (sample C). In this case, the QD height is the same – 2 nm, so differences in emitter properties between the two samples can be neglected. According to calculations (Section 3.3), these two InP shells provide distinctive regimes – with 55 nm shell being too thin to support any guided mode, while 150 nm shell forming a proper single-mode waveguide.

The QD position with respect to the substrate is about 500 nm for all the samples. Since the NW total height is correlated with shell thickness, height value spans from 1300 nm for sample B to 3200 nm and 3500 nm for samples A and C, respectively. The average NW spatial density for sample A and C is below 1 NWs/μm$^2$ (estimated based on several SEM images). Although it would, in principle, enable single QD spectroscopy, the total sensitivity of our current setup is not high enough to measure time resolved PL signal from a single QD-NW in this case (Section 3.4). Therefore, all measurements were intentionally performed on QD-NW ensembles (few thousand of QD-NWs), also to enhance the optical response. Since NWs from sample B do not support waveguiding of the optical mode along the NW, the PL collection efficiency is expected to be much lower. For this reason, in order to have a sufficient level of optical signal, the spatial density increased to 8.1$\pm$0.2/μm$^2$ for this sample.

*Table 1. Overview of the main parameters of investigated samples*

| Sample | QD height | NW shell thickness | NW spatial density |
|---|---|---|---|
| A | 10 nm | 150 nm | 0.7$\pm$0.3 /μm$^2$ |
| B | 2 nm | 55 nm | 8.1$\pm$0.2 /μm$^2$ |
| C | 2 nm | 150 nm | 0.42$\pm$0.04 /μm$^2$ |

### *2.2 Electronic structure calculations*

For theoretical calculations of the electronic structure, the QD is modeled with realistic geometry and composition based on structural measurements of the grown structures, including material intermixing at the interfaces.

The lattice mismatch between InAs and InP causes strain, which considerably modifies the energy spectrum of the QD. To account for this effect and get the resulting strain tensor field, we used the continuous elasticity approach [28], where the elastic energy functional is minimized numerically. Then, the piezoelectric field is calculated taking the polarization up to the second order in strain into account [33], with the parameters from Caro et al. [34]. All this serves as the starting point for the subsequent wavefunction calculations. The electron and hole single-particle states are obtained using the eight-band k·p model [34], [37]. The implementation details are widely described in Refs. [38], [39]. The k·p model material parameters come from fitting the $sp^3d^5s^*$ TB band structures of Jancu et al. [40] and from Vurgaftman et al. [41].

The exciton states are found within the configuration-interaction model, where we took a basis of the 20 lowest electron and hole states. The n-th exciton state is then given by:

$$|X_{\mathrm{n}}\rangle = \sum_{i=1}^{20}\sum_{j=1}^{20} c_{ij}^{(n)} a_i^{\dagger} h_j^{\dagger} \ |vac.\rangle$$

where $c_{ij}^{(n)}$ are complex coefficients resulting from the diagonalization of the Hamiltonian matrix, |vac.> is the vacuum state, $a_i^{\dagger}$ and $h_j^{\dagger}$ are the creation operators for the single-particle electron and hole states, respectively.

The oscillator strength of the transition between the n-th exciton state and the vacuum state is then given by:

$$f_n = \frac{2}{m_0 E_n} \sum_{\mu=x,y,z} \sum_{i,j}^{20} \left| c_{ji}^{(n)} \left(p_{\mu}\right)_{ij} \right|^2 (1)$$

where $E_n$ is the exciton energy, and $(p_{\mu})_{ij}$ is the momentum matrix element between the i-th electron and j-th hole state, calculated using the Feynmann-Hellman theorem [42]. Finally, the radiative lifetimes are calculated from [43], [44]:

$$\tau_n = \frac{6\pi\epsilon_0 m_0 c^3 \hbar^2}{n_r\left(\frac{E_n}{\hbar}\right) E_n^2 f_n} (2)$$

where $n_r(\omega)$ is a frequency-dependent refractive index for InP.

### *2.3 Photonic simulations*

The calculations of the QD-NW photonic properties to evaluate the influence of Purcell effect on the lifetime of excitonic states are performed using finite-difference time-domain (FDTD) method implemented in commercial Ansys Lumerical software [42]. The method is based on solving Maxwell equations in time and space domains in complex structure geometries. To calculate how the NW width affects the Purcell factor ($F_p$), the InP NW is first modelled for a square cross-section, infinite length and no taper. This model allows us to study the fundamental mode confinement without focusing on NW tapering details that might introduce additional reflections. The square cross section is assumed as approximation based on the results of structural characterization of the grown samples (see Fig. 1(a)). The quantum dot is simulated as a point-size electric dipole, placed at NW's center and polarized perpendicularly to the NW vertical axis. Based on the measured PL spectra of the samples differing in InP shell thickness, the central emission wavelength of 1200 nm is chosen for this part of the simulations. In order to study the dispersion of $F_p$, to be directly compared with samples with the same NW shell thickness but emitting at different wavelengths, a broad-band simulation is performed with a realistic NW model. The NW is placed on an InP substrate and the NW length, cross-section width, taper angle and dipole position values are taken from structural data of the samples A and C. The NW width at the position of the emitter in simulations is 350 nm, as the realistic NW stem has the width defined by the Au colloid size to be around 50 nm and the surrounding shell is 150 nm thick, in agreement with the structural data. The NW length is 3500 nm and the dipole is placed 500 nm above the substrate on NW axis. The gold catalyst at the tip of the NW has subwavelength size and is located far away from the dipole, thus it is not included in the model as its influence on the results has been verified to be negligible. In all the simulations, absorbing perfectly matched layer boundary conditions are set for the computational domain.

### *2.4 Optical spectroscopy*

The optical measurements are all performed at 10 K. In the photoluminescence setup, a continuous-wave 640 nm laser is employed for non-resonant optical excitation. Due to the low signal level resulting from thin NW shell in sample B (no waveguide mode supported and Purcell inhibition expected) the excitation power in this case is higher (55 μW) than for samples A and C (25 μW). The laser beam is focused by a lens onto the sample surface. The laser spot diameter is approx. 100 μm, which is big enough to record PL spectra from QD-NW ensemble of over 3 000 NWs at once for each sample. The sample is mounted on a cold finger of a closed-cycle He refrigerator. The detection system consists of a 1 m focal length monochromator equipped with a 150 grooves/mm grating blazed for 1200 nm and a nitrogen cooled InGaAs linear array detector which translates into the practical spectral resolution of 50 μeV.

Time-resolved PL measurements are performed with non-resonant excitation, using a 805 nm pulsed laser with pulse duration of 50 ps and repetition rate of 20 MHz, 40 MHz or 80 MHz, depending on the PL decay time to make sure that the whole the decay curve falls between the excitation pulses. The excitation power is the same as in the PL experiment and is chosen to be high enough to record reliable data but lowest possible to reduce the influence of processes affecting the radiative lifetime (see the discussion below). After being spectrally-resolved by the above-mentioned spectrometer, the emission is coupled to a single mode optical fibre leading to a superconducting nanowire single photon detector, operating with >80% efficiency at the telecom O- and C-band, with maximum dark counts rate below 100 cps and timing jitter below 30 ps.

## 3 Results

### 3.1 *Structural characterization*

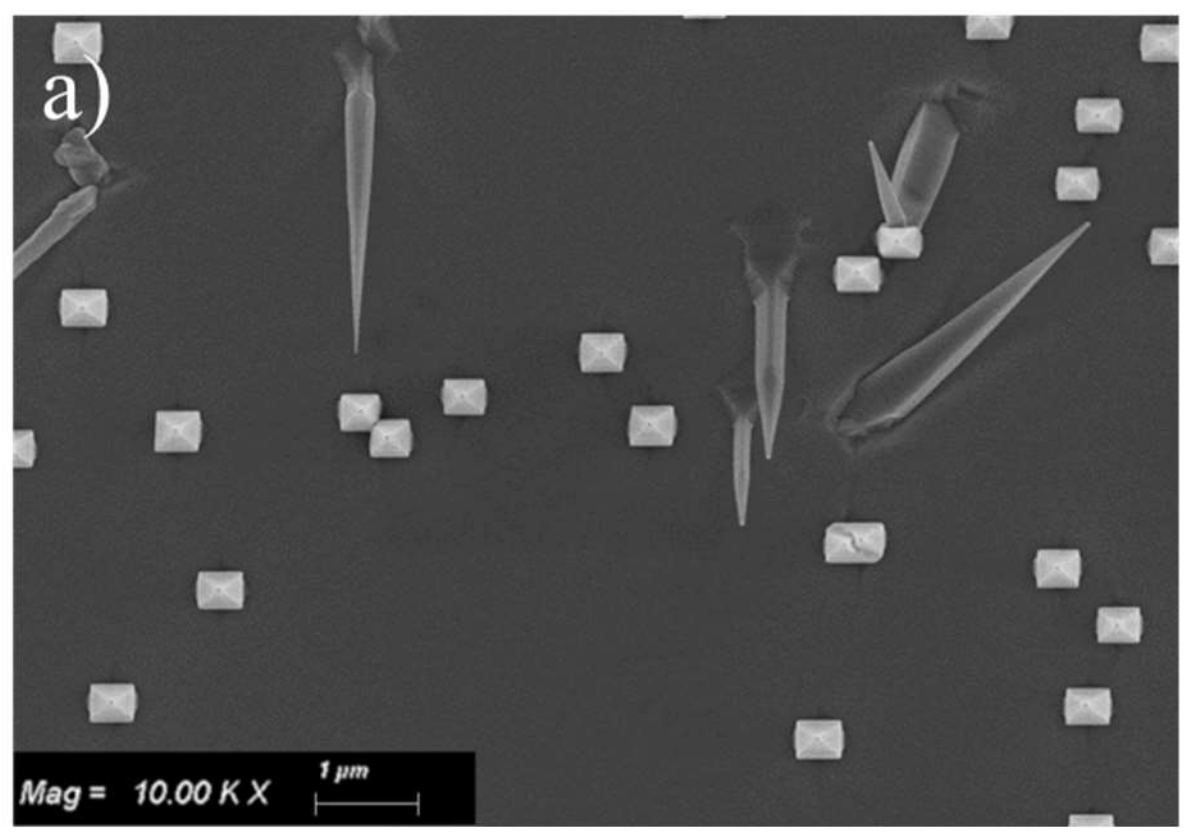

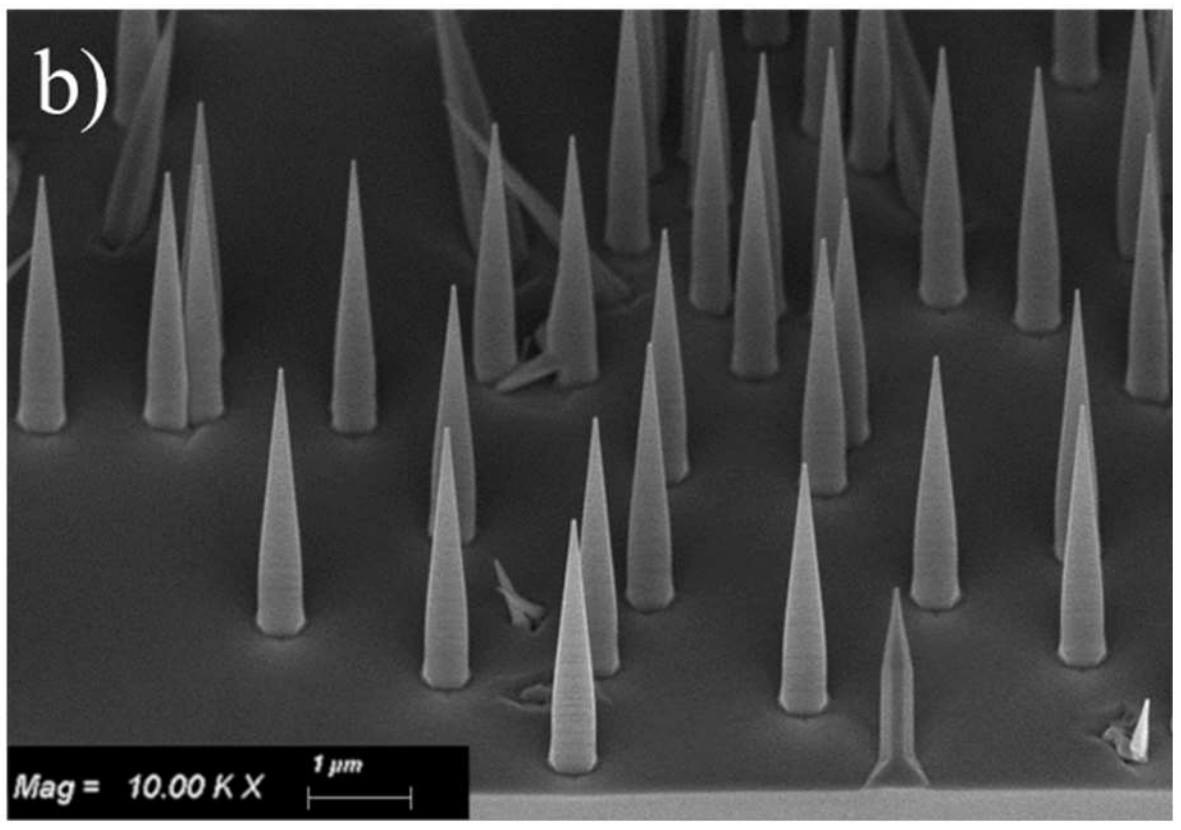

*Fig. 1. SEM images of ZB QD-NWs of sample C from the top view (a) and from 45° tilted view (b)*

Examples of SEM images from the top view and the 45° tilted view are presented in Fig. 1(a-b). The NWs have a square-like cross-section, which is characteristic for the zincblende InP NWs grown in these conditions [21]. There is no significant NW surface roughness, nor visible defects. In order to study the QD structural quality, the HRSTEM measurements are performed on a sample analogous to sample A, but grown without the step of radial shell growth and with a thinner InP tip above the QD. This leads to only a ~4 nm thick InP passivation layer between the QD and the surrounding vacuum. Fig. 2(a) presents a STEM image of the nominally 10 nm high QD. In general, the QD planar size is determined by the Au catalyst size. Since the gold nanoparticle shrinks during the QD growth [21], the QD has a shape of a truncated pyramid, with an average base and top widths of 50 and 45 nm, respectively. The QD height is confirmed to be 10 nm by STEM and EDS measurements. During QD axial growth, a simultaneous InAsP radial growth results in a 2 nm shell "dripping" from the QD edges, as observed by EDS measurements (Fig. 2(a-b)). The existence of such an additional, unwanted ultra-thin InAsP, being a side-effect of the growth mode, has already been reported by us, and it is present only in the case of high QDs growth. For 2-4 nm QDs, the shell is not formed [21]. Fig. 2(b) shows an EDS map of atomic composition of arsenic and phosphorus. Indium has not been included for a clearer data analysis. Axial and radial profiles of the QD are shown in Fig. 2(c-d), respectively. These values have been calculated from the raw data extracted from Fig. 2(b) and corrected by subtracting the InP shell contribution to the signal (see Supplementary Material for more details). The images show sharp interfaces within the EDS mapping resolution in both radial and axial direction (~2 nm). To get some statistics, the structural measurements were performed on 3 QD-NWs from this sample. The QD size and composition turned out to be reasonably uniform, with QD heights close to the nominal value (between 10 and 12 nm) and average composition of $InAs_{0.72}P_{0.28}$.

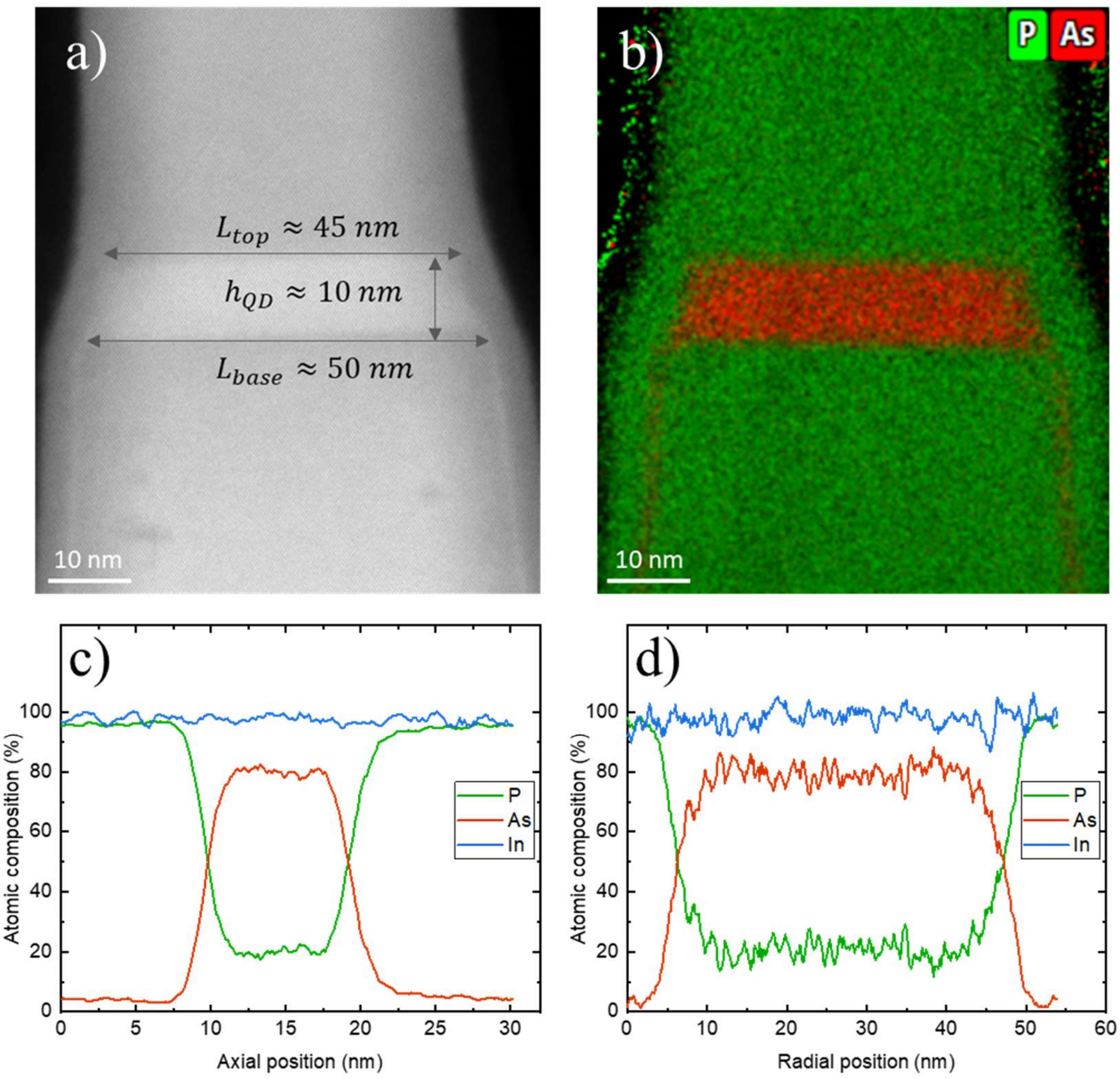

*Fig.2 HRSTEM image of the QD with determined dimensions (a) EDS map of the same NW QD (b) As, P and In atomic distribution as a function of the QD height (axial direction) (c) and width (radial direction) (d)*

### *3.2 Calculated excitonic properties*

We performed theoretical calculations, whose aim is to find correlations between QD parameters and radiative lifetime. It is known that the radiative recombination depends on QD details, such as size, shape and composition, which influence the overlap between electron and hole wavefunctions and the relation between exciton Bohr radius and QD size. The model follows the realistic QD geometry based on structural characterization, with the QD shape modeled as a truncated pyramid defined by its height *h*, as well as the base ($L_{base}$) and the top lengths ($L_{top}$). In the calculations, we take $InAs_{c(r)}P_{1-c(r)}$ material distribution inside the quantum dot, given by Gaussians [43]:

$$c(\overline{r}) = c_{min} + (c_{max} - c_{min}) \exp\left\{-\frac{(x^2 + y^2)}{2\,\sigma_r^2}\right\} \exp\left\{-\frac{z^2}{2\,\sigma_z^2}\right\}$$

Following the EDS composition profiles which suggest some material intermixing between QD and the surrounding InP matrix, we also assume an additional gaussian blur of composition with the standard deviation of 0.6 nm [43]. It is important to notice that the EDS profiles provide information integrated over the direction perpendicular to the imaged plane. One should also keep in mind that there is some distribution of QD parameters within the QD ensemble, so in the results of optical experiments ensemble averaging takes place, which might hinder resolving the more subtle effects. Therefore, the calculations were performed as a function of the QD parameters, with the strongest influence on the QD emission energy and carrier lifetime, namely QD height and As composition. The point **r** = (0,0,0) corresponds to the center of the QD. If not stated otherwise, we take h = 10 nm, $L_{base}$ = 50 nm, $L_{top}$ = 45 nm, $c_{max}$ = 0.7, $c_{min}$ = 0.75 $c_{max}$, $\sigma_r$ = 15 nm, and $\sigma_z = \frac{\sqrt{2}}{3}h$. The resulting exemplary composition profile is shown in Fig.3

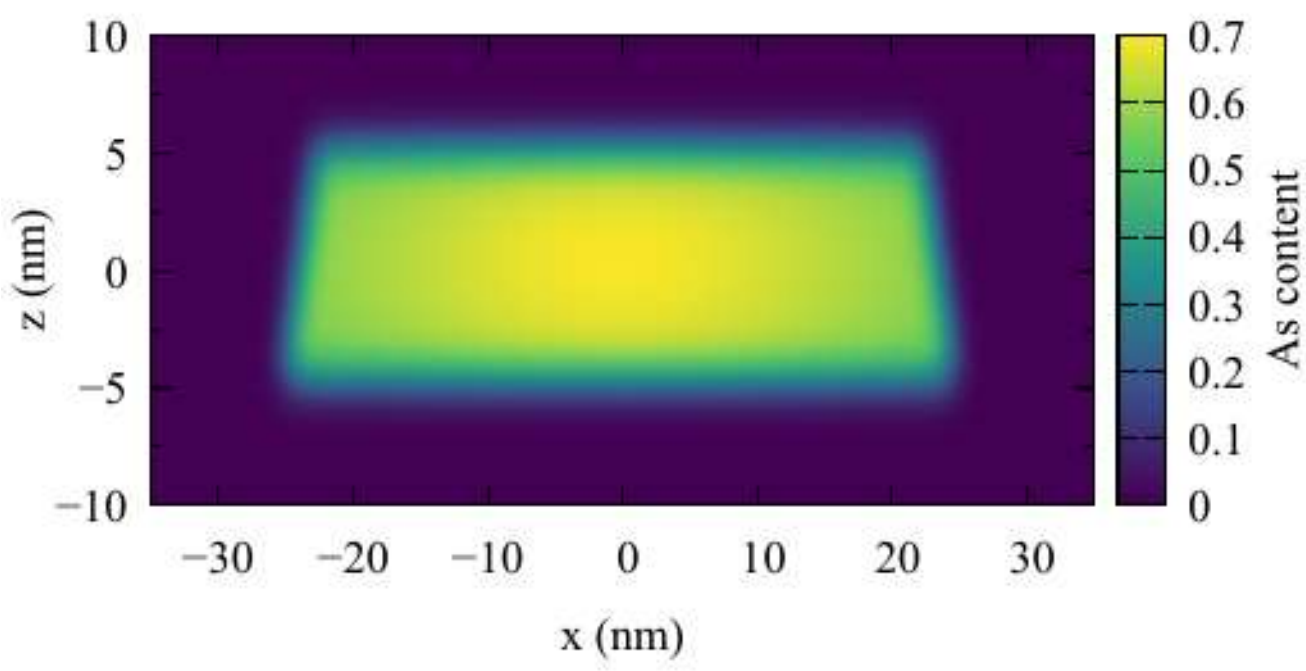

*Fig. 3. Composition profile of $InAs_xP_{1-x}$ QD in InP NW with As content colour-coded adapted in the calculation of strain and further in numerical 8 band k·p simulations.*

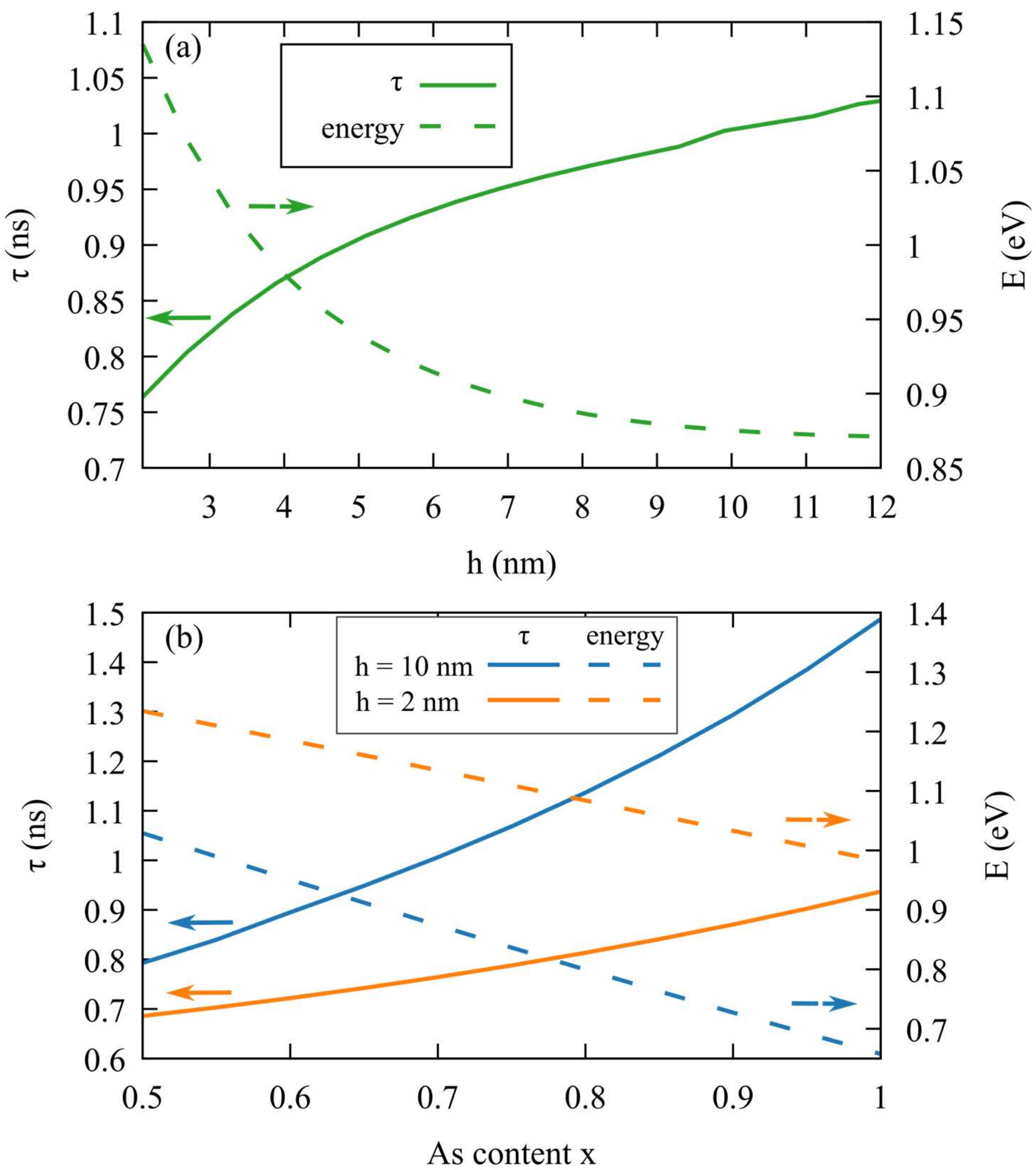

*Fig. 4. The lifetimes (solid lines, the scale on the left) and the energies (dashed lines, the scale on the right) of the exciton ground state as a function of the QD height (a) and the maximum* $c_{max}$ *composition for two distinct QD heights of 10 nm (blue) and 2 nm (orange) (b).*

With the established QD model, we calculate the exciton ground-state energy as a function of the QD height (see Fig. 4(a)). As expected, increasing the QD height lowers the exciton energy due to quantum size effect. The corresponding radiative lifetime increases with growing QD height. This dependence arises from the interplay of several factors. Substituting Eq. (1) into Eq. (2) gives the scaling of the lifetime with the exciton energy approximately $\tau_n \propto \frac{1}{E_n}$. However, the final dependence is affected by the other properties of the system. As demonstrated in Figs. 5(a-b) below, for higher QDs, the electron–hole wavefunction overlap increases, resulting in larger oscillator strengths (due to effective-mass-driven differences in the electron and hole wavefunctions leakage into barriers, similarly as for narrow vs wide quantum wells). The results are also strongly modified by correlations.

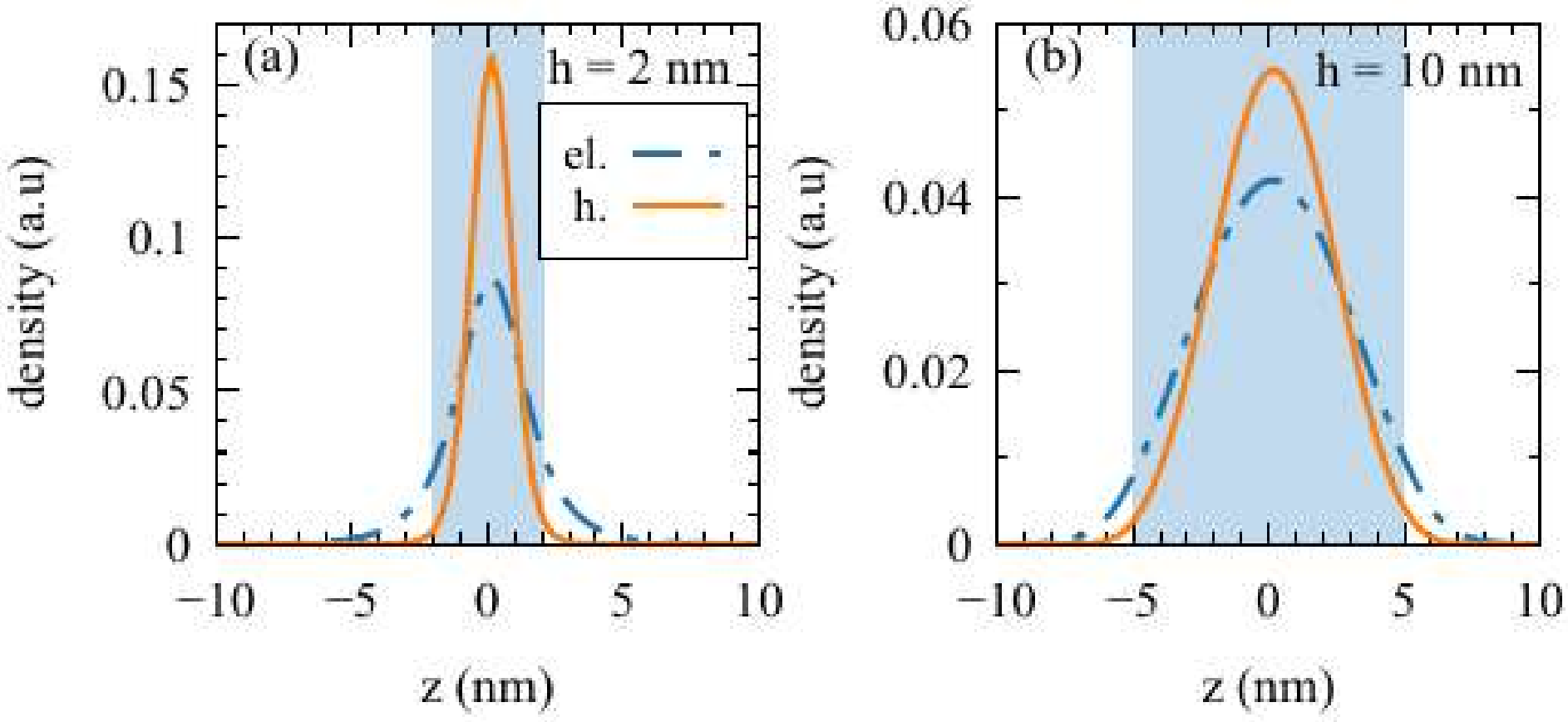

*Fig. 5. The electron and hole probability density (the ground states) in the growth direction for h = 2 nm (a) and h = 10 nm (b) with shaded region marking the QD size.*

We perform analogous calculations for the exciton energy and lifetime, yet for the dependence on the As content. We vary the $c_{max}$ value, while keeping the ratio $c_{min} = 0.75c_{max}$. As shown in Fig. 4(b), the ground-state exciton energy decreases approximately linearly with increasing arsenic content. This behavior can be attributed to the relatively small bowing [41] of the energy gap $E_g$ in $InAs_xP_{1-x}$ alloy and the fact that the optical transition energy is mainly driven by the changes of the QD material energy gap. One can see that the results for two QD heights (h = 2 nm and h = 10 nm) differ in the line slopes. This can be linked to the fact that varying As content modifies the effective masses (which for the electron is 0.074 for InP and 0.024 for InAs). Since changes in the effective mass alter the strength of quantum confinement, it is manifested in the resulting energies. This is the second factor affecting the transition energy dependence on QD composition, due to effectively different confinement for smaller and larger dot for the same range of effective masses.

The calculated lifetimes grow with the increasing arsenic content, and their dependences come from an interplay of several factors, which include: $\tau_n \propto \frac{1}{E_n^2}$ scaling, the changing carriers' probability distributions, and the impact of correlations. We also note that the calculated lifetime is indeed strongly affected by the Coulomb correlations, which enhance the oscillator strength. For example, at h = 10 nm and $c_{max} = 0.7$, the calculated lifetime is 1.01 ns in the full model but 1.47 ns without the correlations (the Hartree approximation). For h = 2 nm and $c_{max} = 0.7$ it is 0.76 ns and 1.42 ns, respectively. This indicates stronger correlation effects compared to the standard InGaAs/GaAs QDs case [39], related to much larger lateral sizes of the investigated dots (50 vs 20 nm).

*3.3 FDTD simulation*

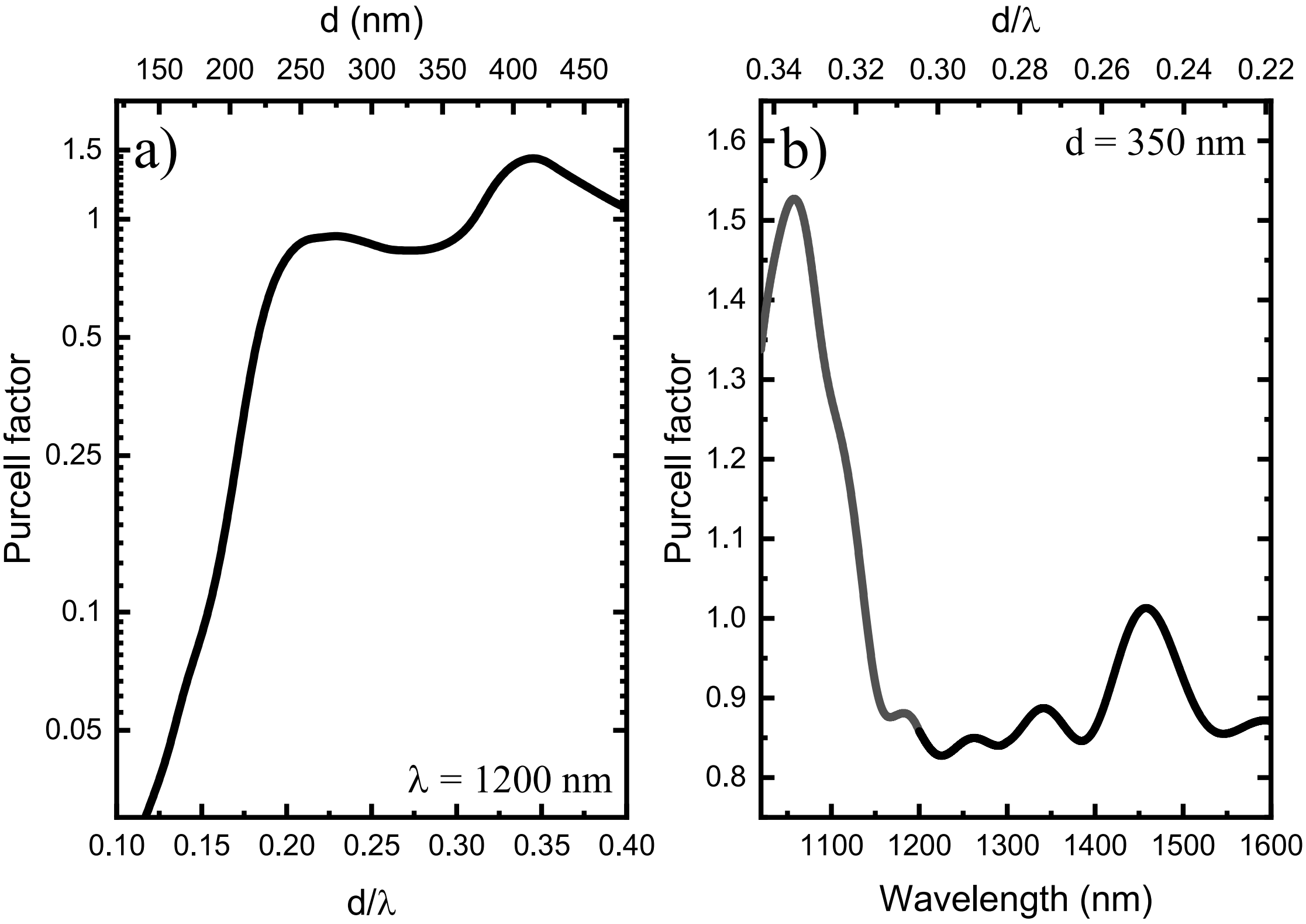

*Fig. 6 Purcell factor ($F_p$) as a function of infinitely long, untapered NW width for 1200 nm dipole emission wavelength (a) $F_p$ as a function of emission wavelength for sample A NW geometry parameters (b)*

The theoretical k·p model allows us to calculate the radiative lifetime based on QD wavefunctions overlap and Coulomb correlations, which is dependent on the QD characteristics. Now we study how this lifetime is modified by the QD photonic environment, without focusing on the emitter details by performing FDTD simulations to calculate the $F_p$. In particular, such calculations allow estimating the difference in the radiative lifetime between samples B and C, resulting from emission inhibition in thin NWs, as well as between samples A and C, resulting from change in the mode confinement regime because of different emission wavelengths.

Fig. 6(a) presents numerically calculated $F_p$ in an infinitely long, untapered InP nanowire for different values of the NW width d. For the ratio of NW width to dipole emission wavelength between 0.20 and 0.31, the dipole emission couples to the single fundamental guided $HE_{11}$ mode, keeping the $F_p$ roughly constant, close to 1. For $d/\lambda>0.31$, the $F_p$ rises, as the second optical mode $HE_{21}$ becomes confined in the NW, effectively increasing the local density of states to which the dipole can emit. For $d/\lambda <0.20$, the NW is too thin to confine the $HE_{11}$ mode, so the emission becomes dominated by the non-guided "leaky" radiative modes. Thus, we observe a rapid decline in $F_p$ and consequently spontaneous emission inhibition.

In Fig. 6(b), the $F_p$ as a function of emission wavelength is presented with the QD-NW geometry of sample A. The $F_p$ is closest to 1 when the fundamental $HE_{11}$ mode is optimally confined and remains $F_p$>0.8 in a wide spectral range. At short wavelengths, the d/λ ratio becomes big enough to support the $HE_{21}$ mode. When the NW d = 160 nm, as in the case of sample B, the $F_p$ remains small, with values <0.1 in the 1100-1600 nm range.

The results show that, based on the relative $F_p$ ratio, the change in radiative lifetimes between the samples A and C should be small, no more than 20%, as the NW with 350 nm shell provides broad enough bandwidth for QDs emitting in both the short and long wavelengths. However, for sample B we expect an at last 10-fold increase in radiative lifetime, caused by the photonic environment influence. While experimentally measured decay curves are affected by both the QD exciton oscillator strength and the NW-geometry-driven photonic confinement in case of sample B, the emission inhibition due to poor NW waveguiding is the dominant effect, affecting the observed elongated lifetimes (see the next section).

### *3.4 Measured emission dynamics*

In order to experimentally verify the calculations, optical measurements are performed on the grown samples. The dispersion of experimentally averaged emission lifetimes measured on QD-NW ensemble is presented for each sample in Fig. 7(a-c), on top of the PL spectra measured at the same arbitrarily chosen spot on the sample surface, corresponding to some QD-NW sub-ensemble. The inset of Fig. 7(d) shows an example of normalized PL decay curve for each sample, registered at various emission wavelengths. For all the curves the rise time is below the setup resolution of ~80 ps and the mono-exponential fit function is used to determine the photoluminescence decay times.

Given the distribution of As content between 69% and 76%, the theoretically calculated emission wavelength for samples B and C, having 2 nm QDs, is in the range of 1080-1140 nm. Sample A, having slightly lower As content and 10 nm QDs is expected to emit between 1390 nm and 1530 nm (Fig. 4). These values are well reflected in the experiment. For sample A, the emission range is even broader, reaching 1580 nm (fully covering the application relevant 3$^{rd}$ telecom window), which might suggest slightly broader QD size distribution. The emission from samples B and C is centred around 1120 nm and 1175 nm, respectively. While the spectral range of sample B is in perfect agreement with calculations, emission from sample C is close to the results from modelling, but redshifted. The slight shift might be caused by differences in QD thickness and/or composition, since these two samples were grown at different times.

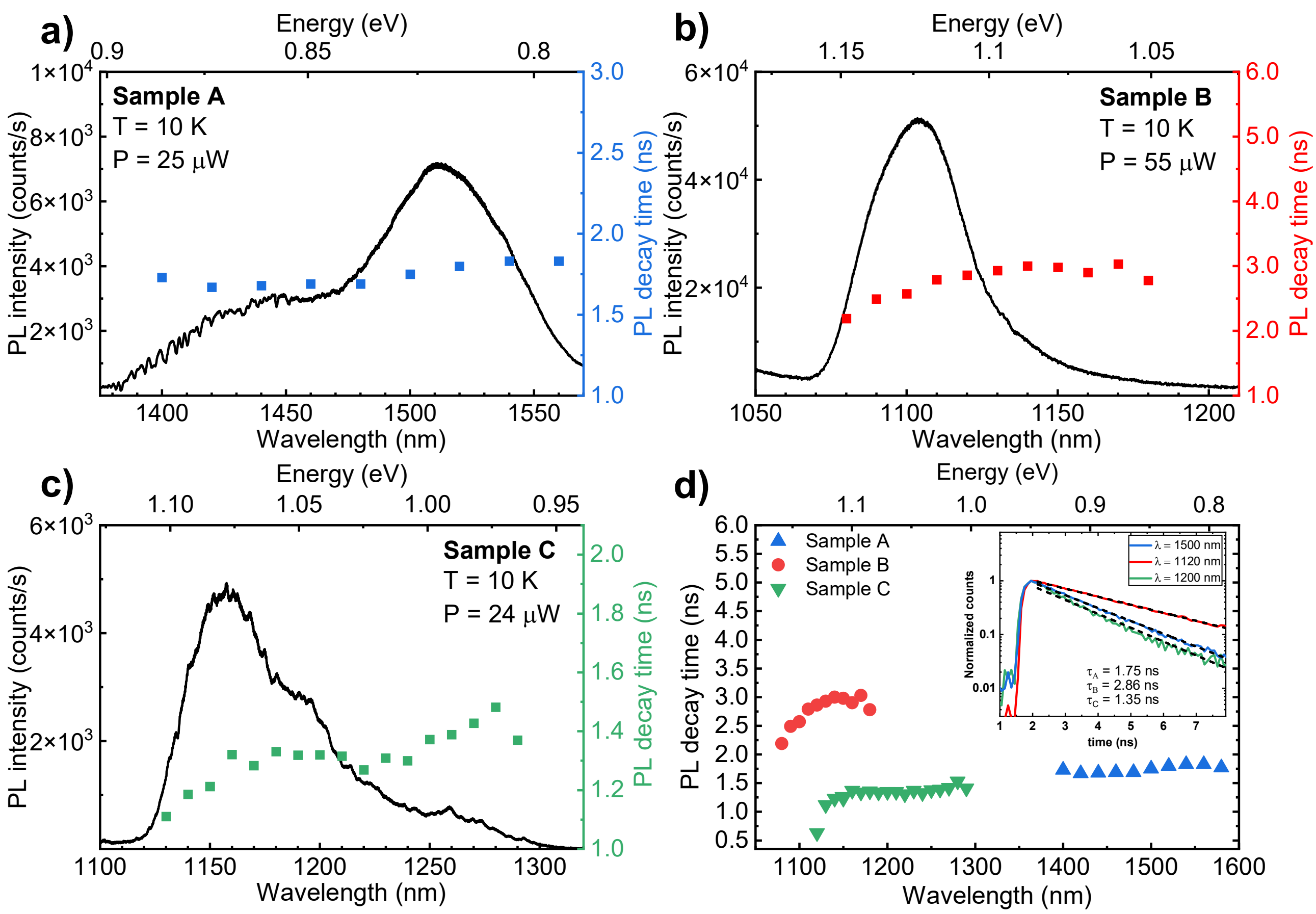

*Fig.7. PL spectra of QD-NW ensembles and decay time dispersion within each sample measured at 10 K (a-c) PL decay time dispersion comparison for all studied samples. The graph inset presents exemplary fitted decay curves (d)*

The collective comparison of PL decay lifetimes for all the samples is shown in Fig. 7(d). Sample A exhibits decay times with an average of ~1.8 ns. The value is bigger than ~1.35 ns measured for sample C, which has analogous NW shell but smaller QDs. The difference is influenced by two factors, as predicted by the theory: radiative decay time lifetime decrease with decreasing As content, but also increase with increasing QD height. However, the difference in the As content is only a few percentage points, which makes this effect weaker than the increase in QD height from 2 to 10 nm. According to the theoretical calculations (Fig. 4), the first effect changes the radiative lifetime by 0.05 ns, while the second one by more than 0.25 ns. The measured PL decay times for both the samples are also slightly elongated by the Purcell factor values below 1, with the effect more pronounced for sample C.

While the decay time for sample C has an average value of 1.35 ns, sample B, having nominally the same QD but thin NW shell not supporting any guided mode, exhibits more than twice longer decay times, with an average of 2.8 ns. This is qualitatively in agreement with numerical photonic calculations shown earlier, which demonstrated Purcell inhibition as the NW width becomes too thin to confine the fundamental guided mode. Quantitatively, the increase is smaller than in our simulations and in the literature [23], [30]. The main difference is that in both cases the lifetime is determined for a single QD or a few QDs in a single NW [23]. Here, we average the lifetime over an ensemble of thousands of QD-NWs, obviously with stronger influence of emitters having shorter lifetimes. What is more, the contributions come from various excitonic complexes and excited states, which usually exhibit shorter lifetimes than the ground state exciton.

## 4 Conclusions

In summary, we investigated both theoretically and experimentally, the factors that influence the emission and exciton dynamics in ZB InAsP/InP QD-NWs emitting in the near infrared spectral range. Structural characterization enables determining QD shape, size and composition crucial for the QD model. The obtained PL decay times of QD-NWs with thick shell vary from 1.1 ns to over 1.8 ns, when the emission wavelength changes from 1100 to 1600 nm, with longer lifetimes observed for higher QDs, with higher As content, which emit in the application relevant 3$^{rd}$ telecommunication window. This can be traced back to the inverse dependence of the radiative recombination time on the emission energy, competing with increased electron-hole wavefunction overlap for larger QDs. Arsenic content mostly influences the confinement regime by changing its depth (by the band gap dependence) and due to lower carrier effective mass in InAs and hence larger exciton Bohr radius, as compared to InP. Photonic environment is responsible for the difference in lifetimes between the samples differing in the shell thickness, as also obtained from calculations suggesting radiative process inhibition by one order of magnitude for thin NWs due to Purcell effect. This effect seems to be alleviated by the fact that the measurements are performed on the QD-NW ensemble. The measured inhibition is smaller for the thin shell sample, which might also be caused by stronger contribution of the higher energy states to the emission (as stronger excitation was necessary in this case). What is worth mentioning, the measured decay times for QD-NWs with thick shell are relatively short compared to literature data for InAsP/InP QD-NWs emitting in the telecom spectral range, where in all the cases they exceed 2 ns [18], [19], [20], [31]. The lower value translates to a higher emission rate, which is beneficial for fast photon generation, but might also suggest increased contribution from non-radiative recombination. The effect depends on structural material quality and presence of carrier traps in the QD vicinity and usually can be minimized by experimental conditions (low temperature and low excitation). However, the decay curves characteristic (monoexponential decay) and short lifetimes obtained from the calculations for such dots do not suggest significant non-radiative contribution.

## Acknowledgements

The authors acknowledge funding from European Innovation Council (EIC) Pathfinder Open program under Grant Agreement No. 101185617 (QCEED) and the PNRR MUR project PE0000023-NQSTI. We acknowledge the financial support provided by the European COST Action OPERA (CA-20116) through the short-term scientific mission program. The authors also acknowledge the Center for Instrument Sharing of the University of Pisa (CISUP) for the TEM facilities. This work has been also supported by the project "Transmission Electron Microscope with EELS and EFTEM System for the Structural and Compositional Characterization of Materials and Devices for 5G+" (TSI-064100-2023-31), funded by the Spanish Ministry of Economy, Trade and Enterprise through the UNICO I+D 6G Programme (2023 Call), within the framework of the Recovery, Transformation and Resilience Plan (PRTR), financed by the European Union – NextGenerationEU. Created using resources provided by Wroclaw Centre for Networking and Supercomputing (http://wcss.pl).